\documentstyle [12pt,twoside]{article}
  
\begin{document}
 
\centerline{\large{\bf Nonstationary Stochastic Resonance}\normalsize} 

\vspace*{1cm}

\centerline{\bf Redouane Fakir}
\vspace*{.5cm}
\centerline{\em Peter Wall Institute for Advanced Studies}
\centerline{\em Cosmology Group, Department of Physics \& Astronomy}
\centerline{\em Department of Interdisciplinary Studies}  
\centerline{\em University of British Columbia, 
6224 Agriculture Road, Vancouver, B. C. V6T 1Z1, CANADA}
\vspace*{1.cm}
\centerline{(Physical Review E, in press)}
\vspace*{2.cm}
{\bf
It is by now established that, remarkably, the addition of noise
to a nonlinear system may sometimes facilitate, rather than hamper
the detection of weak signals. This phenomenon, usually referred to as
stochastic resonance, was originally associated with strictly periodic
signals, but it was eventually shown to occur for stationary
aperiodic signals as well. However, in several situations of
practical interest, the signal can be
markedly nonstationary. We demonstrate that the phenomenon of 
stochastic resonance extends to nonstationary signals as well,
and thus could be relevant to a wider class of biological and
electronic applications. Building on both nondynamic and
aperiodic stochastic resonance, our scheme is based on a multilevel
trigger mechanism, which could be realized as a parallel 
network of differentiated threshold sensors. We find that
optimal detection is reached for a number of thresholds of
order ten, and that little is gained by going much beyond that number.
We raise the question of whether this is related to the fact that
evolution has favored some fixed numbers of precisely this order of magnitude
in certain aspects of sensory perception.
}
\clearpage

Random fluctuations are ubiquitous in physical systems,
be it electronic circuits, sensory neuron networks, or
any number of a large set of situations of great physical
interest. This is what made the discovery of stochastic
resonance$^{1-3}$ potentially a milestone in the field
of signal detection. Since the early 80's, an increasing 
number of authors have identified or suspected the
occurrence of stochastic resonance in a wide variety of
contexts$^{4}$, from paleo-climatology$^{1,5}$ to biological sensory
systems$^{6-8}$ to nonlinear electronic circuits$^{9-10}$.

In the course of research on how a newly identified effect of
galactic gravitational waves$^{11}$ could be detected, it was 
pointed out that stochastic resonance could be relevant in 
that context$^{12}$, since the effect in question involved
(electromagnetic) {\em noise} in a fundamental way. It was subsequently
found that applying two simple numerical
threshold triggers to the data could allow the extraction of the gravitational
signal from the noise$^{13}$. In fact, this was but a specific realization 
of more general work on stochastic resonance then recently published$^{14-16}$.
The particularly interesting finding had just been made, that systems
as simple as a nondynamic trigger mechanism could display all the
desirable features of stochastic resonance. The implication was that
stochastic resonance, although depending crucially on the nonlinearity of the
dynamics involved, may not necessarily depend on the {\em details} of that nonlinearity. Thus, stochastic resonance could be a more widespread phenomenon than previously thought.

More recently still, stochastic resonance was demonstrated 
for aperiodic signals$^{17-24}$. More specifically, it was shown
that stochastic resonance can arise for a signal that is 
a weak arbitrary fluctuation around some constant value, in other words,
a weak stationary signal. This was an
important first step towards showing that stochastic resonance 
is actually relevant to real-life signals, which rarely come
in exact sine waves.  

Here, we attempt to push these findings further by showing that,
through stochastic resonance, very simple (biological or artificial)
devices can also detect sub-threshold signals that are not
stationary. Many realistic situations call for the detection of signals
of just such a nature. In fact, it is not always easy in nonlinear
electronic circuits to stabilize a signal enough that it can be considered
as rigorously stationary, while in biological applications the signal
that is sought is seldom truly stationary. We are essentially interested
in the case where the aperiodic signals already considered in the
literature$^{17-24}$ have an additional nonstationary structure,
i.e. a nontrivial overall time profile. It is this low frequency
aspect that we focus on, and we do not attempt to reproduce the
known properties of (higher-frequency) aperiodic stochastic resonance. 

Our detector is a  multilevel trigger system. This could be realized,
for example, as a number $N$ of single threshold sensors mounted in 
parallel. Stochastic resonance in such cooperative systems has been shown
to exhibit very attractive features$^{20,25,26}$, such as not requiring a finely
tuned amount of noise to accomplish the detection$^{20}$. Our system here
is the equivalent of a network of parallel {\em differentiated} sensors:
We consider a trigger mechanism characterized by a number $N$ of thresholds of
different heights. As was noted in the literature$^{20}$,
such a threshold differentiation can do little to improve the detection
of {\em stationary} signals. We now show that such multilevel trigger
systems can improve the detection of signals that are
markedly nonstationary. We first sketch an analytical account of
roughly how one expects the system to behave, and then present the 
results of the actual numerical simulation in Fig.s 1-4. We have consigned
part of the discussion of the simulations to the captions of these figures.

Consider first the usual {\em single}-threshold trigger system$^{14-16}$.
Each time the input exceeds the threshold value $B_{1}$, the system
generates a pulse of height $H$ and width $W$. The response is thus
a pulse train $P(t)$, where $t$ indicates the time. If the input is
Gaussian white noise with rms $\sigma$ and high-frequency cutoff $f_{0}$,
then the response averages to 
\newline\begin{equation}
<P> = H W {f_{0}\over\sqrt{3}} \exp{(-B_{1}^{2}/2\sigma^{2})}  \  \  .
\end{equation}\newline
Now, a sub-threshold signal $S(t)$, varying on time scales much larger
than $1/f_{0}$, is added to the input. The rate of threshold crossing
will now vary in time, and the resulting response $R(t)$ can be 
approximated by the smeared response $\tilde{R}(t)$ (average
response over time intervals that are longer than the noise correlation time
and shorter than the signal's characteristic time) can be approximated
by$^{16}$
\newline\begin{equation}
\tilde{R}(t) = H W {f_{0}\over\sqrt{3}} \exp{(-(B_{1}-S(t))^{2}/2\sigma^{2})}  \  \  .
\end{equation}\newline

Moreover, in realistic situations (especially in biological ones),
there is an upper limit to the {\em effective} pulse firing rate:
A neuron, for example, can potentially fire a maximum of about
$500$ times per second.
(Note, however, that as far as its interaction with the external environment
is concerned, the neuron generally has the equivalent of a moving window of about $0.1$
second, so that it can react to changes in the external stimulus
only at an effective rate of $10$ times a second. Hence, the correct
way to model the problem is highly case dependent.
We are mostly interested here in the detection of a signal that is
buried in {\em external noise}, i.e. in situations where the stimulus
itself is noisy. This is the situation that arises when trying to identify
the presence of a deterministic signature within a stochastic
input. One must thus distinguish between internal and external noise
in a more realistic model.)
 
There is then an effective upper limit on the rate at which the system
can respond to the input, which we model by  allowing
each trigger to fire a maximum of about $1000$ times during its exposure to a one second-long signal. When the input starts crossing that trigger's threshold
at a rate faster than $1000$ Hz, the threshold starts firing at its
maximal rate, contributing a constant value to the total output.
The firing rate's upper limit is typically
much smaller than the frequency cutoff $f_{0}$ of the noise. Multiplying
that limit by $H W$, one obtains an upper ceiling $R_{upper}$
on the value of the response function. This sudden flattening of $R_{r}(t)$,
the more realistic response function, 
can be modeled algebraically in different ways, depending on the internal
dynamics of the particular system considered. However, the effect of this
plateau in $R_{r}(t)$ that  we
are interested in here can be captured by the following
simple model:
\newline\begin{equation}
R_{r}(t) = min \{ R_{upper} , R(t) \} \   \  .
\end{equation}\newline

We want to study the correlation of this output function $R_{r}(t)$ with the 
injected signal $S(t)$, which should determine whether the latter can 
be detected by the trigger system. We follow the literature$^{17,18}$ in using
the zero-lag correlation coefficient, with the modification that our signal
does not average to zero (Fig.1a). In our numerical 
simulations (Fig.s 1-4), we of course correlate directly with the raw (pulse-wise) 
response $R_{r}(t)$, not with its smeared version $R_{r}(t)$. 
But the smearing approximation is sufficient for capturing
analytically the main new features brought about by non-stationarity. Note
also that in several applications, especially in biological ones, it is
often the gross time profiles of the nonstationary signal and of the response
that are relevant,
i.e. their smeared versions. Thus, we define
\newline\begin{equation}
C = {< ( R_{r}(t) - <R_{r}(t)> ) ( S(t) - <S(t)> ) >
\over 
< ( R_{r}(t) - <R_{r}(t)> )^{2}>^{1/2} 
< ( S(t) - <S(t)> )^{2}>^{1/2} }  \   \   ,
\end{equation}\newline
and, to gain some insight into the behavior of this correlation 
coefficient, we briefly consider the behavior of its ``smeared''
counterpart
\newline\begin{equation}
\tilde{C} = {< ( \tilde{R}_{r}(t) - <\tilde{R}_{r}(t)> ) ( S(t) - <S(t)> ) >
\over 
< ( \tilde{R}_{r}(t) - <\tilde{R}_{r}(t)> )^{2}>^{1/2} 
< ( S(t) - <S(t)> )^{2}>^{1/2} }  \   \   .
\end{equation}\newline

If the total integration time is $\Delta t$, we shall compare 
the slope of the time profiles against the 
factor (see Fig.1)
\newline\begin{equation}
a \equiv { B_{1}\over \Delta t} \  \  ,
\end{equation}\newline
which is the slope of the diagonal across the area underneath the
threshold $B_{1}$ in Fig.s 1a,b. For example, a roughly linear
signal which, when smeared, has a slope $x$ that is much smaller than $a$,
can be considered as virtually stationary. 
In that particular case, by injecting the right amount of
noise to the system, it is possible to make the total input
(signal plus noise) hover just about the threshold for the duration of the
integration. The time features of the threshold-crossing rate will then
partially reflect those of the signal, and the latter could be detected 
along the lines of (stationary) aperiodic stochastic resonance$^{17-24}$.

A rather different situation arises if the slope $x$ is of the
order of $a$, i.e., if the signal is markedly nonstationary. Consider then
the simplest 
nonstationary case, that of a signal the smeared value of which grows
almost linearly with time, e.g. $S(t) = x t + y$, where $x$ and $y$ are
quasi-constants. If we also want the signal to start from $S(t=0)=0$
and still remain subthreshold at the end of integration time 
($t=\Delta t$), we can choose $y=0$ and we need $x$ to be somewhat smaller 
than (though still of the same order as) the non-stationarity factor $a$.
Note finally that a more general nonstationary time profile could be
qualitatively well approximated by a succession of linear sections such as
the one above.

If some noise is now added to this signal, 
say with an rms $\sigma \sim B_{1}/3$ ,
then the total input is first well below the threshold $B_{1}$, then well above it, with,
somewhere in between, a brief critical time when the total input is just
about the threshold, as in the stationary case. The output is hence first
essentially zero. Then it quickly grows, over the critical time, and reaches 
a saturation value where it stabilizes for the remainder of the integration.
(See Fig.2a.)
This saturation can be roughly traced analytically in the behavior of
equations (2,3), upon substituting for $S(t)$ the  
linear form $x t$: If we define the normalized time 
\newline\begin{equation}
\tau \equiv t/\Delta t  \  \ ,
\end{equation}\newline
which grows from $0$ to $1$ over the integration time, we find
\newline\begin{equation}
R_{r}(\tau) = min \left\{ \ R_{upper}\  , \ H W {f_{0}\over\sqrt{3}}
\exp{\left( {-B_{1}^{2}\over 2\sigma^{2}} 
\left[ 1 - {x^{2}\over a^{2}}\tau (2{a\over x} - \tau) \right] 
\right)} \ \right\} \  \  .
\end{equation}\newline
Because the signal is below the threshold, we have $\tau < a/x$ in equation (7).
On the other hand, we are considering the strong nonstationary
case, so $a/x \sim 1$. In addition, we have, by definition, $0<\tau <1$.
Equation (7) then implies the following behavior of the system, which is confirmed by direct numerical simulation in Fig.2.
Consider first the case where $R_{upper}$ is arbitrarily high. Then
equation (7) implies that
$R_{r}(\tau)$ grows almost linearly until it approaches its maximum
($\sim H W f_{0}$) 
at $\tau = 1$, and hence the correlation with the input signal is
quite high. This has the implication that the single-threshold systems
used so far in (stationary) aperiodic stochastic resonance$^{17-24}$ have
also the potential to detect nonstationary signals under certain
circumstances, namely when the pulse firing rate can be extremely fast$^{28}$.

However, in the more generic case where $R_{upper} << H W f_{0}$, 
the realistic response $R_{r}(\tau)$ is not so well correlated
to the signal, as it is a function that flattens quickly when the
total input crosses above the threshold. This is where multiple,
differentiated thresholds can substantially improve the detection.
Let us then mount,  in parallel with the first, some additional
trigger systems, in all identical to the first except for their  
progressively higher thresholds (Fig.1): $B_{1} < B_{2} < ... < B_{N}$.

The second threshold will then saturate at a higher value of $\tau$
and its output, when added to output from the first threshold, will
resemble a two-step function. With the $N$ thresholds in parallel, 
one obtains an $N$-step total response function $R_{r}(\tau)$. As $N$ increases
from the order of $1$ to the order of $10$, the $R_{r}(\tau)$ profile
rapidly approaches a roughly linear shape, and one expects the correlation
coefficient of equation (4) to improve notably. However, a further increase
in $N$ towards very large values can be expected to bring about only
a marginal improvement in the correlation, and this is in fact what is
found numerically (Figs.2,3). 

As a short preview of our numerical simulations, we first note that we used a signal that increases almost
 monotonically with time (Fig.1). Again, more arbitrary
nonstationary profiles can often be divided into such quasi monotonic sections. Also, we purposely did not add higher frequency
structure to the signal because (1) we wanted to isolate the features 
brought about by nonstationarity, (2) stochastic resonance has 
already been successfully demonstrated for (stationary) aperiodic
high-frequency signals$^{17-24}$, and (3)
the signal that a system (especially a biological one) attempts to detect
in the nonstationary case is often a rough time profile, and higher 
frequency components are then of little relevance.

In nonstationary situations where both the overall time profile 
{\em and} the higher frequency details of the signal are important,
one can first try to detect that rough (low-frequency) time profile
as described here, then filter out the low frequencies and treat
the residual signal (a {\em stationary} high frequency signal) with
aperiodic stochastic resonance techniques$^{17-24}$. 

Finally, as might be expected from previous nondynamical$^{14-16}$
and aperiodic$^{17-24}$ stochastic resonance studies, our differentiated
multi-threshold system, injected with nonstationary aperiodic signals,
clearly displays stochastic resonance as well (Fig.4), although it
is its potential for the {\em detection} of sub-threshold signals
that is of prime importance to us here. 

Describing now are simulations in more detail, we turn first
to Fig.1a, which shows the 
$N$-level trigger system (here $N=5$) consisting of a barrier $B_{1}$
and $N-1$ higher thresholds equally spaced between $B_{1}$ and a highest
threshold $B_{N}$, chosen here to be three times $B_{1}$. Other ways of distributing the differentiated thresholds are of course possible, but most
do not lead to qualitatively different results. The signal is clearly 
nonstationary, and yet it remains sub-threshold, i.e. undetectable in 
principle, throughout the integration. (The time has been normalized
so that the integration runs from $\tau =0$ to $\tau=1$.)

In Fig.1b, the deterministic input of (a) is replaced by a random looking input, obtained by adding
a large amount of noise to the previous sub-threshold signal. The noise
here is a low-pass filtered, zero-mean Gaussian white noise. The input
now clearly exceeds the barrier $B_{1}$ and a few more thresholds.
When any of the thresholds is exceeded, the system fires a pulse of standard
(but otherwise arbitrary) height $H$ and (narrow) width $W$.
At any given time $\tau$, the response $R(\tau)$ is the sum of the outputs
from all the thresholds exceeded at that time $\tau$. I.e., if $B_{j}$
is the highest threshold exceeded at $\tau$, then $R(\tau)$ is a pulse
of width $W$ and height $j\times H$.

Fig.2a shows the sub-threshold input signal (thin dashed curve) of Fig.1a and
the response $R(\tau)$ of the trigger system. (Both are re-scaled here
and in figures 2b and 2c, for easier visual comparison.) When only one
threshold (that is, $B_{1}$) is activated, the response is roughly
a step function: For small values of $\tau$, most of the points in
Fig.1b are below $B_{1}$ and hence produce a zero response, while
for large values of $\tau$, most of the points are above $B_{1}$ and 
produce a response equal to $1\times HW$ (scaled here to $1$.) 
One can see that this one-step response function is only marginally
correlated to the underlying signal, which translates into a
low value of the correlation coefficient for $N=1$ in Fig.3.

From Fig.2b one can see that when $N$, the number of thresholds, 
is greater than one ($N=10$ in Fig.2b), each of the exceeded thresholds (here, there are four)
contributes a different step function to the response. Hence, the latter
takes the aspect of a multi-step function (again re-scaled here
as in (a)) that follows roughly the nonstationary profile of the signal.
There is a clear improvement of the correlation between the
response and the signal, confirming our earlier heuristic argument
which used the smeared approximation ${\tilde R}(\tau)$ to the response
$R(\tau)$.

Because the quasi step functions due to the individual thresholds are
partially stochastic, they also contribute some broken lines to the 
total response, as can be seen in (a) and (b) above. When, as in Fig.2c,
 the number
of thresholds becomes large ($N=100$ in Fig.2c), the (now very short) steps
and the stochastic broken lines become comparable in length, resulting 
in a scattered response function. As a consequence, the correlation between signal and response is not much stronger than in the
case $N\sim 10$, hence the leveling off of the correlation coefficient
as a function of $N$ in Fig.3.  
 
Fig.3a shows the correlation coefficient of equation (2), which is the value of
the normalized correlation function (between the response and the
signal) at zero-lag. The increase of the correlation with the
number of thresholds levels off at $N\sim 10$, as was foreseen in the
above discussion of Fig.2. (The first three values of $N$
produce the same correlation because, for these small values, the space
between thresholds is large enough that only $B_{1}$ is exceeded.)
Thus, a number of order-of-magnitude 
$10$
of differentiated , individual triggers, mounted in parallel (summed
outputs), is singled out for the optimal detection of a nonstationary
signal.  

Fig.3b confirms that the main feature of Fig.3a, i.e. the plateau of the correlation coefficient
as a function of the number $N$ of thresholds, is robust in the large-$N$
limit. In both Fig.3a and Fig.3b, the calculations was performed
using a noise with rms $\sigma$ equal to half the height of the first
barrier $B_{1}$. 
 
Finally, Fig.4 shows the correlation coefficient $C$ of equation (4) as a function of $\sigma$,
the rms of the noise, divided by the lowest threshold $B_{1}$.
The system clearly displays stochastic resonance. When there is more than
one threshold ($N>1$), the decrease of $C$ becomes slightly sharper when
$\sigma$ exceeds a value of order $1$. This is a consequence of having a fixed highest threshold, namely $B_{1}+2\times B_{1}$ (see Fig.1):
When the noise is large enough that its tip starts exceeding the highest threshold (which occurs when $\sigma\sim B_{1}$), the whole system starts behaving more like
a one-threshold system, hence the sharper decrease in $C$ (compare with
the $N=1$ curve.) Note that for $N$ of the order of $10$ or larger, all curves are nearly identical
to the $N=1000$ curve shown in Fig.4. So, although there is a clear broadening
of the stochastic resonance curves (and hence less fine tuning of the noise
is necessary to detect the weak signal) for large values of $N$, this system is not as free of fine 
tuning as the (non-differentiated) multi-threshold system described in the (stationary) aperiodic stochastic resonance literature$^{20,25,26}$ referred to earlier in the text.

In all, our study suggests that stochastic resonance might
be exploited for the detection of even strongly nonstationary
sub-threshold signals . 
 We find that a differentiated multiple
threshold system might, in such cases, improve on the performance
of single threshold systems. But this is by no means a universal
conclusion: we find numerically the amount of improvement to be strongly
case dependent (i.e. depending on the shape of the signal, the
noise characteristics, etc.) The results reported here are for
cases where the multi-threshold scheme presents a clear advantage.
A thorough investigation of the comparative performance of differentiated
multi-threshold systems throughout parameter space is underway$^{27}$.
The capacity of {\em modified} single-threshold systems to detect
realistic nonstationary signals has also been investigated recently$^{28}$.

One example of a differentiated multiple threshold system that might
function similarly to our theoretical system is the auditory sensory system
in mammals$^{29}$.  In this system, each primary afferent neuron (from the spiral
ganglion) has its lowest threshold to a particular frequency of sound,  
called its "characteristic frequency." Its threshold is increasingly higher
for sound frequencies farther from the characteristic frequency.  When the
system is stimulated by a weak pure tone of a given frequency, only neurons
with a characteristic frequency near the stimulating frequency, and thus
having low thresholds for those sounds, will fire. As the tone increases in
amplitude, the first-firing neurons will quickly saturate but other
neurons, whose thresholds for the stimulating frequency are higher because
their characteristic frequencies are farther from it, will begin to fire as
their thresholds are exceeded.  In this way, the number of neurons firing
will track the intensity of the pure tone.  If these neurons converged upon
an integrator neuron whose firing depended directly on the firing of
several spiral ganglion neurons with different characteristic frequencies,
the integrator would fire proportionately to signal amplitude.  This system
would give a similar response for a subthreshold signal that
was amplified above the thresholds of the auditory receptors by
environmental or internal noise, i.e., in the stochastic resonance
situation.  Such circuits in the auditory system have been argued to be
responsible for primary coding of auditory signal intensity$^{30}$ and for
integration of intensity information from very brief auditory signals$^{31}$.
 
Our simulations suggested that only a relatively small number of
cooperative trigger mechanisms (of order $10$ in our study) might be
necessary to achieve near-optimal detection of subthreshold signals, at
least nonstationary ones.  This result is reminiscent of the fact that some
key components of certain biological systems consist of a fixed number of
receptors mounted in cooperative neural networks similar to our
differentiated multi-threshold system.   For example, under optimal
conditions a minimum of six photons must each stimulate a different rod in
the retina of the eye, the signals from the six rods being integrated in
the response of a single retinal ganglion cell$^{32}$.  This is not a stochastic
resonance situation, since the rods are roughly identical near-ideal
detectors responding to single photons, and Poisson noise from the source
only degrades that function.  However, it is interesting that evolution has
developed such a sparse network that can be so exquisitely sensitive to
environmental signals.  Perhaps there is a more general principle of
diminishing returns operating that keeps such cooperative networks small
even though they are of diverse function.  We speculate that this principle
might be important in the functioning of stochastic resonance in some
biological systems and that evolution may have exploited it to enhance the
detection of a wide class of fitness-relevant environmental signals, in
particular nonstationary signals as described here. 

Clearly, however, much more detailed
studies of biologically (or electronically) plausible multiple-component, stochastic resonance systems, injected with a variety
of realistic signals, must be conducted before such an evolutionary
role of stochastic resonance is confirmed.

\vspace*{2.cm}
\centerline{\bf Acknowledgements}
\vspace*{0.5cm}
I am grateful to L.M. Ward for helping me become better acquainted with
neural processes, as well as for his insightful comments on the reported
results. I would also like to thank P.E. Greenwood for reviewing the paper
and making several suggestions.
I am indebted to W.G. Unruh for being the first
to bring the phenomenon of stochastic resonance to my attention, and to
B. Bergersen for helping me familiarize myself with previous literature
on stochastic resonance. I have also benefited from  
extensive logistical support by the General Relativity
\& Cosmology Group in the Department of Physics, University of
British Columbia, by the Department of Interdisciplinary Studies
at UBC and by T.E. Vassar during the preparation of this paper. 
This work was supported by the PWIAS of UBC and NSERC of Canada.

\clearpage
\hspace*{4.cm} {\bf\large REFERENCES}
\newline\newline\newline
1. Benzi R., Sutera S. \& Vulpiani A, {\em J. Phys.} A{\bf 14},
L453 (1981).
\newline\newline
2. Nicolis C., {\em Tellus} {\bf 34}, 1 (1982).
\newline\newline
3. Benzi R., Parisi G., Sutera A \& Vulpiani A., {\em Tellus} {\bf 34},
10 (1982).
\newline\newline
4. Wiesenfeld K. \& Moss F., {\em Nature} {\bf 373}, 33 (1995).
\newline\newline
5. Nicolis C., {\em J. Stat. Phys.} {\bf 70}, 3 (1993).
\newline\newline
6. Longtin A., Bulsara A. \& Moss F, {\em Phys. Rev. Lett.}{\bf 67},
656 (1991).
\newline\newline
7. Douglass J.K., Wilkens L., Pantazelou E. \& Moss F.,
{\em Nature} {\bf 365}, 337 (1993).
\newline\newline
8. Bezroukov S.M. \& Vodyanoy I., {\em Nature} {\bf 378}, 362 (1995).
\newline\newline
9. Fauve S. \& Heslot F., {\em Phys. Lett.} A{\bf 97}, 5 (1983).
\newline\newline
10. Mantegna R.N. \& Spagnolo B., {\em Phys. Rev.} E{\bf 49} R1792 (1994).
\newline\newline
11. Fakir R., {\em Int. J. Mod. Phys.} D{\bf 6}, 49 (1997);
{\em Phys. Rev.} D {\bf 50}, 3795 (1994); {\em Astrophys. J.} {\bf 426},
74 (1994).
\newline\newline
12. Unruh W.G., private communication (1995).
\newline\newline
13. Fakir R., {\em ``A macroscopic gravity wave effect;''} Los Alamos Physics Archives astro-ph/9601127.
\newline\newline
14. Jung P., {\em Phys. Rev.} E{\bf 50}, 2513 (1994);
{\em Phys. Lett.} A{\bf 207}, 93 (1995).
\newline\newline
15. Wiesenfeld K., Pierson D., Pantazelou E., Dames C. \& Moss F.,
{\em Phys. Rev. Lett.} {\bf 72}, 2125 (1994).
\newline\newline
16. Gingl Z., Kiss L.B. \& Moss F., {\em Europhys. Lett.} {\bf 29},
191 (1995).
\newline\newline
17. Collins J.J., Chow C.C. \& Imhoff T.T., {\em Phys. Rev.}
E{\bf 52}, R3321 (1995).
\newline\newline
18. Collins J.J., Chow C.C., Capela A.C. \& Imhoff T.T., {\em Phys. Rev.}
E{\bf 54}, 5575 (1996).
\newline\newline
19. Collins J.J., Imhoff T.T. \& Grigg P., {\em J. of Neurophysiology}
{\bf 76}, 642 (1996).
\newline\newline
20. Collins J.J., Chow C.C. \& Imhoff T.T., {\em Nature} {\bf 376}, 236 (1995).
\newline\newline
21. Levin J.E. \& Miller J.P., {\em Nature} {\bf 380}, 165 (1996).
\newline\newline
22. Heneghan C., Chow C.C., Collins J.J., Imhoff T.T.,
Lowen S.B. \& Teich M.C., {\em Phys. Rev.} E{\bf 54}, R2228 (1996).
\newline\newline
23. Gailey P.C., Neiman A., Collins J.J. \& Moss F., 
{\em Phys. Rev. Lett.} {\bf 79}, 4701 (1997).
\newline\newline
24. Neiman A., Schimansky-Geier L. \& Moss F., {\em Phys. Rev.} E{\bf 56},
9 (1997).
\newline\newline
25. Pantazelou E., Moss F. \& Chialvo D., in {\em Noise in Physical Systems
and 1/f Fluctuations}, (eds Handel P.H. and Chung A.L.) 549-552 
(American Institute of Physics Press, New York, 1993.)
\newline\newline
26. Moss F. \& Pei X., {\em Nature} {\bf 376}, 221 (1995).
\newline\newline 
27. Fakir R., Ward L.M. \& Greenwood P.E. {\em ``Stochastic resonance in multi-threshold systems;''} in preparation.
\newline\newline
28. Fakir R., {\em ``Nonstationary stochastic resonance in a single
neuron-like system;''} submitted for publication.
\newline\newline
29. I thank L.M. Ward for many details in the following comments about neurons.
\newline\newline
30. Luce R.D. \& Green D.M., {\em Psychological Review} {\bf 79}, 14 (1972).
\newline\newline
31. Ward L.M., {\em Perception \& Psychophysics} {\bf 50}, 117-128 (1991).
\newline\newline
32. Hecht S., Schlaer S. \& Pirenne M.H., {\em J. of General Physiology}
{\bf 25}, 819-840 (1942).

\clearpage
\centerline{\large\bf Figure captions}
\vspace{4.cm}
\centerline{\bf Figure 1}
\vspace{2.cm}
{\em
a) The $N$-level trigger system (here $N=5$) consists of a barrier $B_{1}$
and $N-1$ higher thresholds equally spaced between $B_{1}$ and a highest
threshold $B_{N}$, chosen here to be three times $B_{1}$ 

b) The deterministic input of (a) is replaced by a random looking input, obtained by adding
a large amount of noise to the previous sub-threshold signal. 
}
\clearpage
\vspace{4.cm}
\centerline{\bf Figure 2}
\vspace{2.cm}
{\em
a) The sub-threshold input signal of Fig.1a and
the response $R(\tau)$ of the trigger system, both re-scaled here
and in (b) and (c) below, for easier visual comparison. When only one
threshold (that is, $B_{1}$) is activated, the response is roughly
a step function.

b) When $N$, the number of thresholds, is greater than one (here
$N=10$), the response
takes the aspect of a multi-step function (again re-scaled here
as in (a)) that follows roughly the nonstationary profile of the signal.

c)  When the number
of thresholds becomes large (here, $N=100$), the steps
and the stochastic broken lines become comparable in length, resulting 
in a scattered response function.
}
\clearpage
\vspace{4.cm}
\centerline{\bf Figure 3}
\vspace{2.cm}
{\em
a) The correlation coefficient of equation (4), which is the value of
the normalized correlation function (between the response and the
signal) at zero-lag, as a function of $N$, the number of thresholds.
The increase of the correlation coefficient with $N$ levels off at $N\sim 10$, as was foreseen in the text. Thus, a number of order-of-magnitude $10$
of differentiated , individual triggers, mounted in parallel,
is singled out for the optimal detection of a nonstationary signal.  

b) The result of (a) above, i.e. the plateau of the correlation coefficient
as a function of the number $N$ of thresholds, is robust in the large-$N$
limit. 
}
\clearpage
\vspace{4.cm}
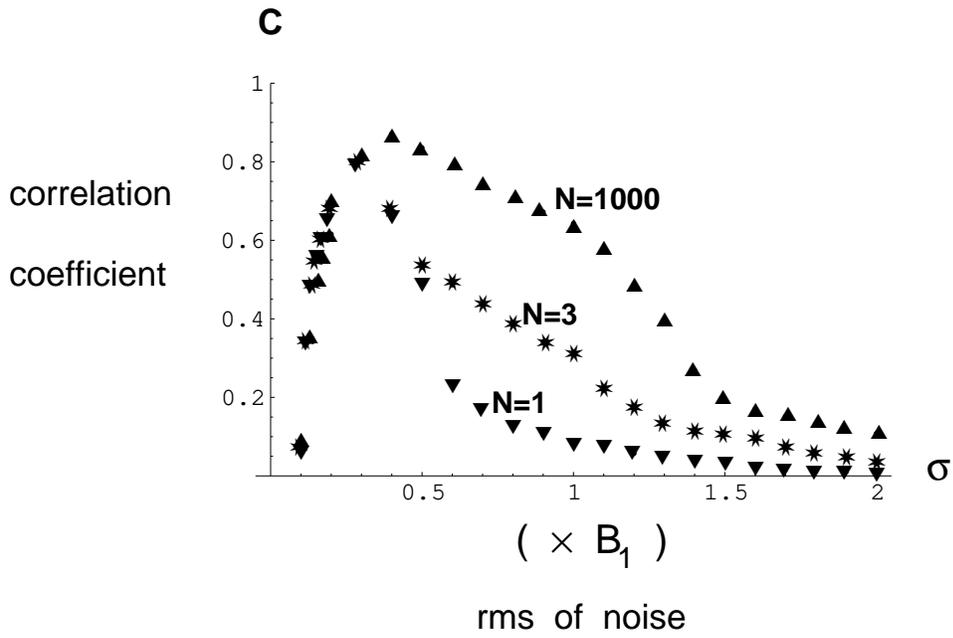
\centerline{\bf Figure 4}
\vspace{2.cm}
{\em
The correlation coefficient $C$ of equation (4) as a function of $\sigma$,
the rms of the noise, divided by the lowest threshold $B_{1}$.
The system clearly displays stochastic resonance. 
(See the text for more on these specific profiles.)
}
 
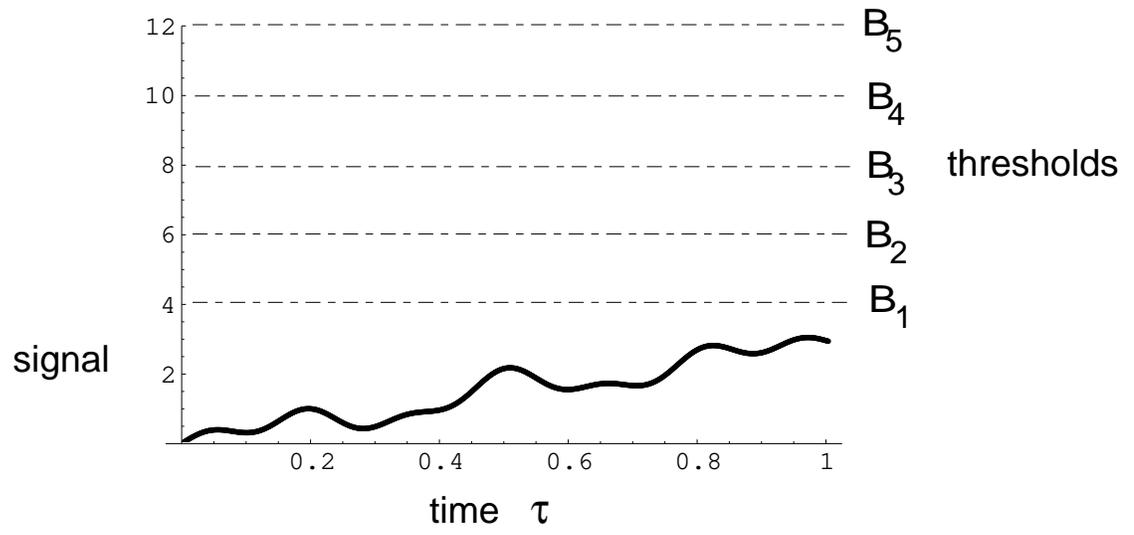
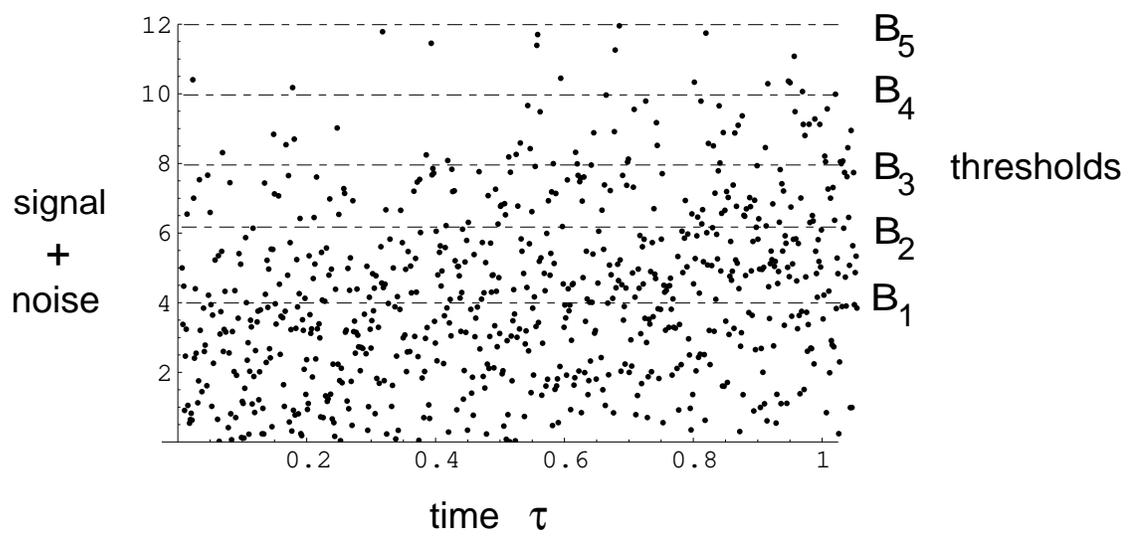
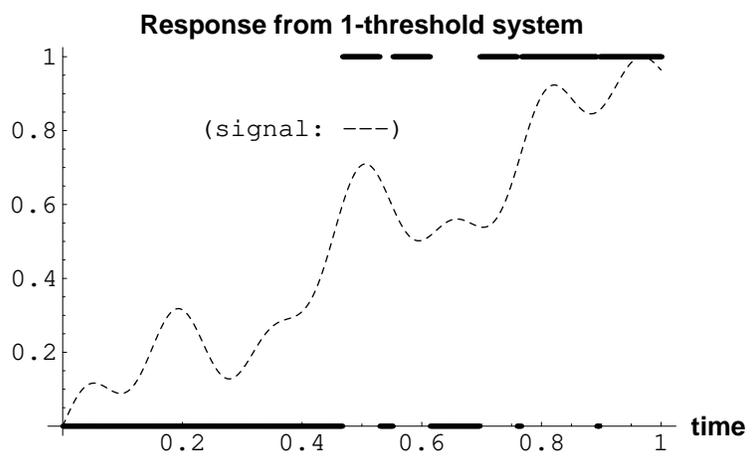
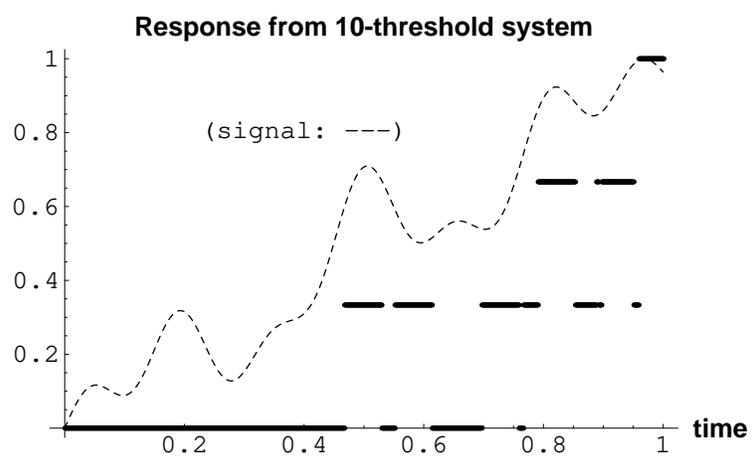
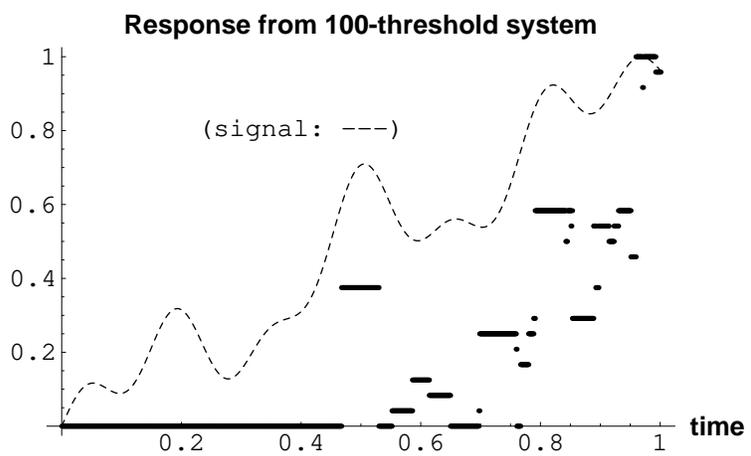
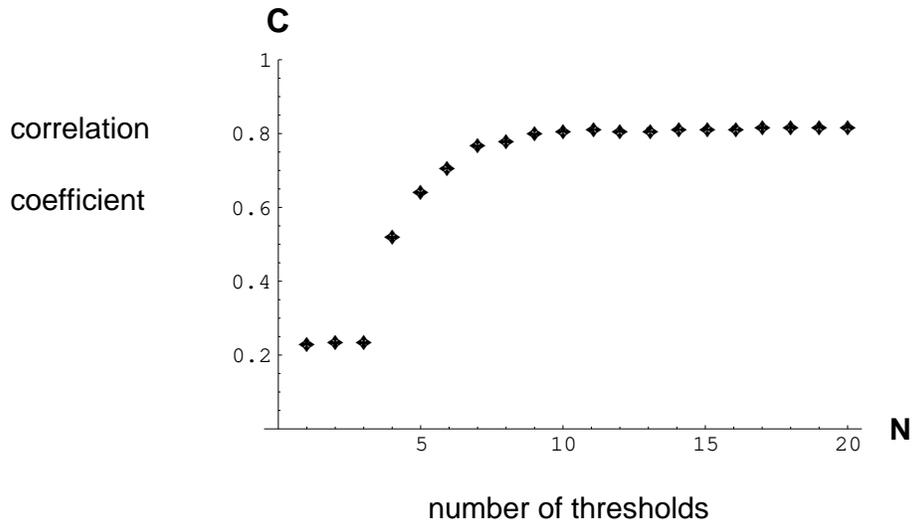
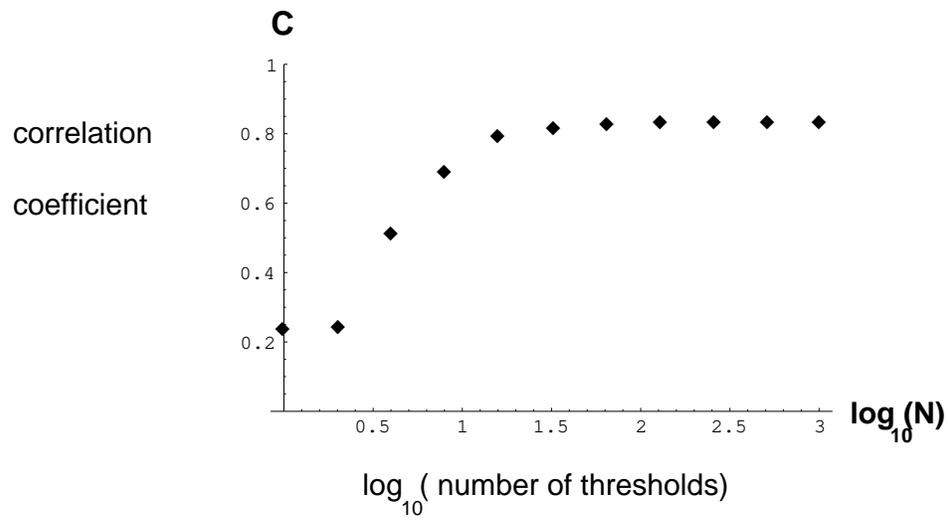
\end{document}